\documentclass{article}
\usepackage[utf8]{inputenc}
\usepackage{hyperref}
\usepackage{emsnewsletter}

\author{Isabel Beckenbach  (FIZ Karlsruhe -- Leibniz Institute for Information Infrastructure, Berlin), Klaus Hulek (Leibniz University Hannover),  
  Olaf Teschke  (FIZ Karlsruhe -- Leibniz Institute for Information Infrastructure, Berlin)
}

\title{The extension of zbMATH Open by arXiv preprints}

\begin{document}

\maketitle

\begin{abstract}  zbMATH Open has started a new feature -- relevant preprints posted at arXiv will also be displayed in the database. In this article we introduce this new feature and the underlying editorial policy. We also describe some of the technical issues involved and discuss the challenges this presents for future developments. 
\end{abstract}

\section{Introduction}

During the last three decades, the arXiv has established itself as the main preprint repository for mathematics. Some years ago, in \cite{MT16a,MT16b}, we analysed the share of publications in the zbMATH corpus available via arXiv versions. There we saw that, despite a very uneven distribution depending on the mathematical subjects, the share of arXiv coverage has been growing constantly. 
Today, the overall arXiv share exceeds a third of all recent mathematical publications, with the figures for several core areas of mathematics being significantly higher, often exceeding 50\%. Moreover, the arXiv share is still growing, although some saturation effects can already be observed in certain areas (e.g., the percentage of about 80\% of publications
in algebraic geometry available on the arXiv has not improved much recently). 
zbMATH Open has adapted its services to this situation, and has been providing links from published documents to available arXiv versions for more than 10 years now. 
Obviously, this required a precise matching algorithm, taking into account the information provided by the arXiv. This information, however, is somewhat unstable. It may very with submission versions, and the citable source strings contain less information than the ones from published journals.
Indeed, our success in matching zbMATH Open entries to arXiv versions has much improved since we started linking to it, and the figures (as evaluated on test sets definable by dois) now indicate 
a very mature state (see Section \ref{tech}).

One obvious advantage is that this ensures the open availability of a significant share of mathematics research -- indeed, despite the growing ecosystem of open and hybrid journals, platforms, and transitory deals, the arXiv accounts by far for the largest share of mathematics publications available through open access \cite{ET20}. 
But the advantages achieved through this matching go much beyond findability and access. 
Namely, this also opens up the path for numerous new investigations and thus new insights. Just to give an example: the links now serve as a proxy to estimate the time between 
submission and publication in various mathematical areas \cite{AT17, AHT17, KT23}. 

Especially the long period between submission and publication in many core mathematical areas, predominantly caused by a thorough peer review process, raises natural questions about the difference 
set: how large is the {\em unpublished} share of the arXiv in mathematics? What are the chances of arXiv submissions to be finally published? And, perhaps most importantly, wouldn’t it be beneficial for a comprehensive information system in mathematics to integrate a well-defined subset of the arXiv into zbMATH Open? In particular, it could serve to improve the visibility of recent work, especially by young researchers whose career chances suffer from long publication delays.
If so, how should one formulate a sound indexing policy that matches closely the zbMATH Open Scope despite the lack of a formal peer review process at the arXiv?

There can be few doubts about the relevance of unpublished arXiv preprints. Not just the share of the platform in recent work, but also the omnipresence of unpublished arXiv submissions in the references of already published documents underline its significance. So the principal decision to enlarge the information available in zbMATH Open by unpublished arXiv preprints was a very natural one. However, the details how exactly to 
realise this in practice, are very involved. In this column we outline the editorial policies and the new features that define the arXiv subset now integrated into zbMATH Open.

\section{The editorial policy}

\subsection{Aims and resulting editorial decisions}

The scope of zbMATH Open, as defined by its editorial policy, has been to index ``all available published and peer-reviewed articles, books, conference proceedings as well as other publication formats pertaining to the scope given above that present a genuinely new point of view" (\url{https://zbmath.org/about/#id_1_1}). This allows different interpretations and is subject to the dynamics of the field and its publication system. Nevertheless, the consistency of this approach has been ensured by the editorial board, the expert staff covering the diverse mathematical subjects, the ongoing maintenance and development of the applied classification system (\url{https://msc2020.org/}), and the supervisory boards.  

Unpublished arXiv preprints per se do not satisfy all conditions of the above definition. 
There is no refereeing process for arXiv, although a mature mediator and endorsement system is installed to ensure both the scientific relevance and formal standards of the accepted submissions. 
The first and foremost condition for integrating unpublished arXiv preprints into zbMATH Open must be that they are clearly distinguishable from published work.
This now happens at two different levels: firstly as a new database type (`arXiv', in addition to Zbl, JFM, and ERAM); secondly as a new document type (`Preprints', in addition to books, and journal and collection articles). Filtering with both facets will currently provide identical results, although this may change by future developments.

\begin{center}
\includegraphics[width=.5\textwidth]{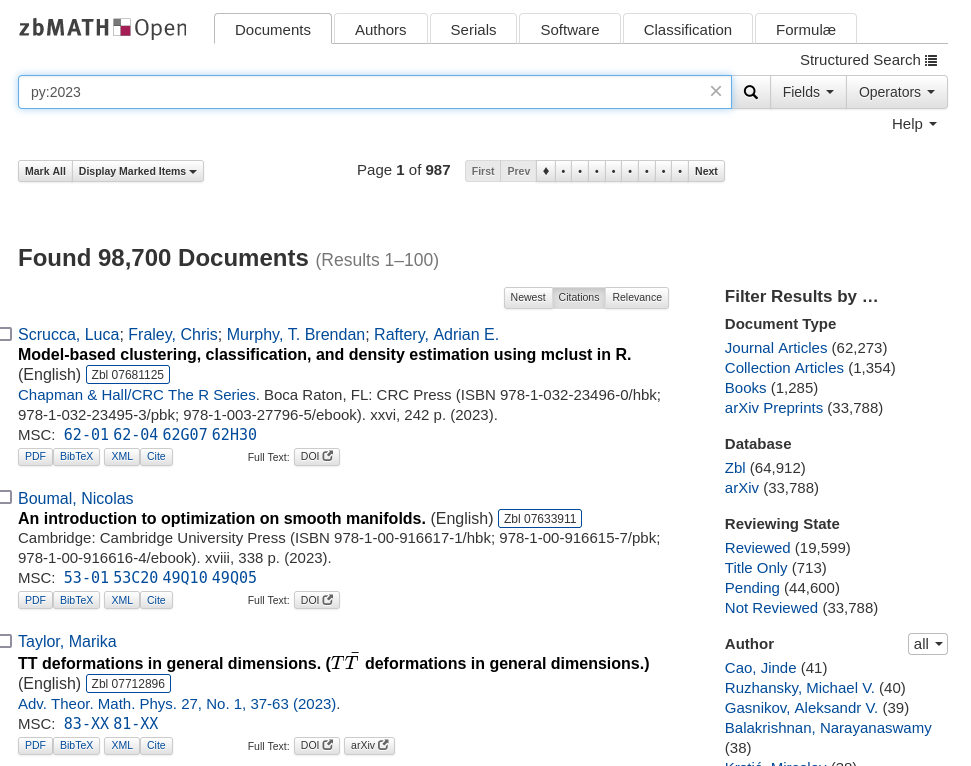}
\captionof{figure}{New filter functions allow to clearly distinguish additional unpublished arXiv entries}
\end{center}

Perhaps the most important distinction from classical zbMATH Open documents is that unpublished arXiv preprints will not be sent out for reviewing. This is due to the very nature of arXiv submissions -- they are not stable versions, and may in fact vary significantly over time. 
To reflect this, they would require multiple reviews, which would both be beyond the resources of zbMATH Open and unsatisfactory for reviewers.
Neither will there be multiple entries for (possibly quite different) arXiv versions, but only one, which is defined by the stable arXiv identifier - along with author and title information, the only metadata required for the entries. The preliminary nature is also reflected by the fact that, contrary to other new zbMATH Open entries, there will be no editorial classification -- only MSCs already provided by authors will be displayed. However, the entries {\em are} integrated into the zbMATH Open author disambiguation, to ensure that they are visible in the author profiles (more details are given below in Section \ref{authors}).

Finally, once more emphasizing the preliminary nature of the arXiv entries, they will be merged with the stable version after publication. After that, the arXiv version will just appear as an additional link, 
precisely as it has been before the arXiv integration.
 
\subsection{Defining the scope}

While this approach aims at following both the perception of arXiv as a preprint repository and the user expectations, it also creates specific challenges to the scope definition. 
There are  several possible errors which may not completely be ruled out, but whose effects need to be minimized. First, of course, matching errors may either create duplicates (in case of false negatives) or missing arXiv entries (for false positive matchings). As explained in Section \ref{matching}, the zbMATH Open arXiv matcher is now sufficiently sophisticated
to ensure that such effects are negligable. However, they cannot be completely ruled out -- e.g. Saharon Shelah (the mathematician with most overall arXiv submissions) has submitted individual book chapters separately to the arXiv, which cannot -- due to the lack of metadata -- be automatically matched to the review of the complete books in zbMATH Open.

Of course, there are also arXiv submissions which will, for a variety of reasons, never be published. In this case, the entries will remain forever in their incomplete form, and eventually form a group of permanent zbMATH Open entries which would not exist otherwise. However, as this reflects the reality of mathematical information, this should be considered an asset rather than a liability.  
It may also happen that arXiv submissions are published in a journal or conference outside the scope of zbMATH Open. In this case, the entry will, somewhat unsatisfactorily, likewise remain forever in its preliminary form -- and, moreover, it may actually not fit very well into the zbMATH Open scope (as evidenced by its final publication venue). We aim to minimize both effects by choosing only a limited and carefully chosen arXiv subset for inclusion.

This may enlarge the risk to err in the other direction: excluding arXiv preprints although their content would be relevant to zbMATH Open. 
Even when they become finally available though publication in a source indexed by zbMATH Open, the information is still missing for a significant length of time, and the authors concerned may feel unduly 
disadvantaged in comparison with other arXiv entries which were includxed. Obviously, there is no ideal solution to this dilemma. The only meaningful option is a thorough analysis of how much in the 
different arXiv categories and subcategories has actually been published within the zbMATH Open scope in the past, and then to define a practical threshold.

The arXiv math category has 32 subcategories currently in use (subcategories that are not in use anymore are irrelevant to the question of which recent unpublished preprints should be included). Of these, 28 have an overlap of more than 60\% with the zbMATH Open corpus, i.e., they have been published in sources indexed in zbMATH Open. Some subcategories even have a share of more than 70\% of the zbMATH corpus, though the differences appear to be more correlated with different publication delays and publication behaviour than with actual scope differences. 

The share for the categories math.GM (General Mathematics), math.HO (History and Overview) und math.IT is significantly lower, although for different reasons. math.GM and math.HO contain a relatively large amount of non-research mathematics (e.g., of educational nature), while math.IT has only been introduced in 2007 as an alias for the subject area information theory  in computer science. For this reason it naturally contains a large number of not primarily mathematical contributions.  

The only borderline case turned out to be math.ST (which is an alias of Statistics Theory, stat.TH) with a share of about 50\%, which mostly reflects the fact that it contains not just mathematical 
research in statistics but is also often cross-re\-fer\-enced from non-mathematical categories with descriptive statistical work. Although it is almost impossible to establish a perfect distinction here, further 
analysis showed that the overlap with the published mathematical research literature increases sufficiently if submissions cross-referenced from non-mathematical arXiv categories are excluded -- so this is the criterion
currently employed as the scope definition for math.ST. On the other hand, the overlap of mathematical physics (math-ph, or as an alias, math.MP) was sufficiently high to be included as a whole.
 
\section{New features for the arXiv entries in zbMATH Open}\label{features}

\subsection{Extent of the newly added information}

Given these premises, slightly more than 200,000 unpublished arXiv preprint have now been added to the database -- a figure which is likely to grow in the future due to the overall publication growth.
But this does not represent a stable corpus in itself, since documents will be both added (through new preprints) and removed (due to their publication). 
About 83,000 documents are (preliminarily) assigned Mathematical Subject Classification numbers as provided by the authors. This figure could be expanded by a rough automated classification based on arXiv subcategories and semantic information, but no final policy decision has yet been taken in this direction. In any case, given the preliminary nature of the documents, there will be no further intellectual classification before publication.

With current figures, combinatorics (about 10,000 items) is the largest area among the newly added preprints, followed by partial differential equations (8,700), number theory (8,600), probability theory (8,300) and algebraic geometry (8,200).

It turned out that also the abstracts could be converted without too many problems, and are thus also displayed along with the entries. This came as a surprise as, in principle, arXiv submission are free in their \LaTeX{} definitions and do not necessarily fit into the zbMATH Open \LaTeX{} framework \cite{ST19}. The inclusion of abstracts allows for more efficient searches and a more informative display for these items. The entries are also integrated into the zbMATH Open author disambiguation (see Section \ref{authors}). On the other hand, there is currently no reference extraction, since the citations formats are highly non-standardized within the arXiv. Moreover, references may also vary significantly between different arXiv versions of the same submission. Therefore, these citations are currently not integrated into the zbMATH Open citation database (but see also the discussion in Section \ref{sec:future}).

\subsection{Features}

As a results, the additional arXiv entries are now, with respect to most of their features, searchable and retrievable. The entries can be filtered by authors, submission year, or MSC subject. 

\begin{center}
\includegraphics[width=.5\textwidth]{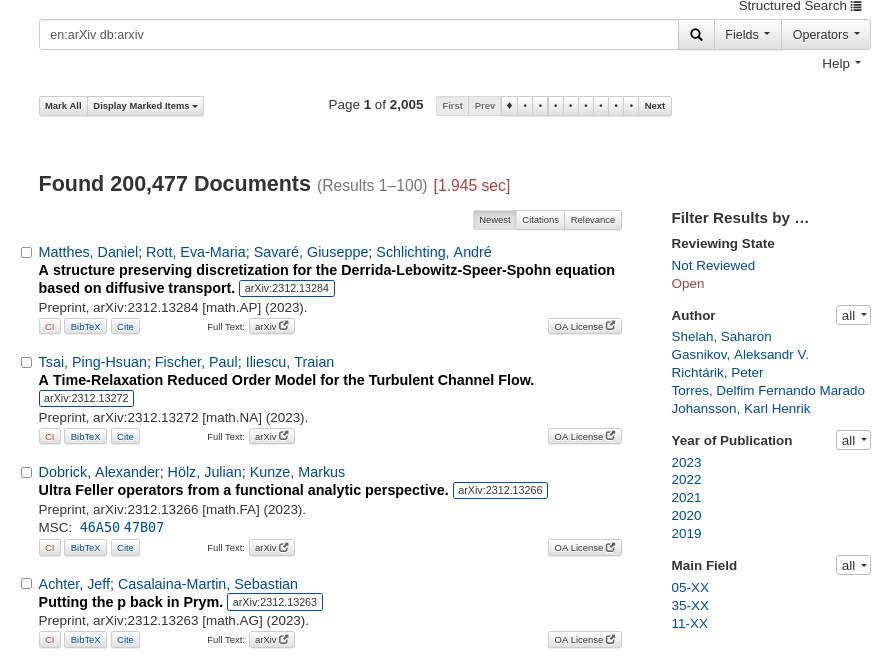}
\captionof{figure}{Filter functions for more than 200,000 newly added arXiv preprints}
\end{center}

Author signatures are linked to profile information, and the additional arXiv submission are included into author profiles, as in the next figure. 
\begin{center}
\includegraphics[width=.5\textwidth]{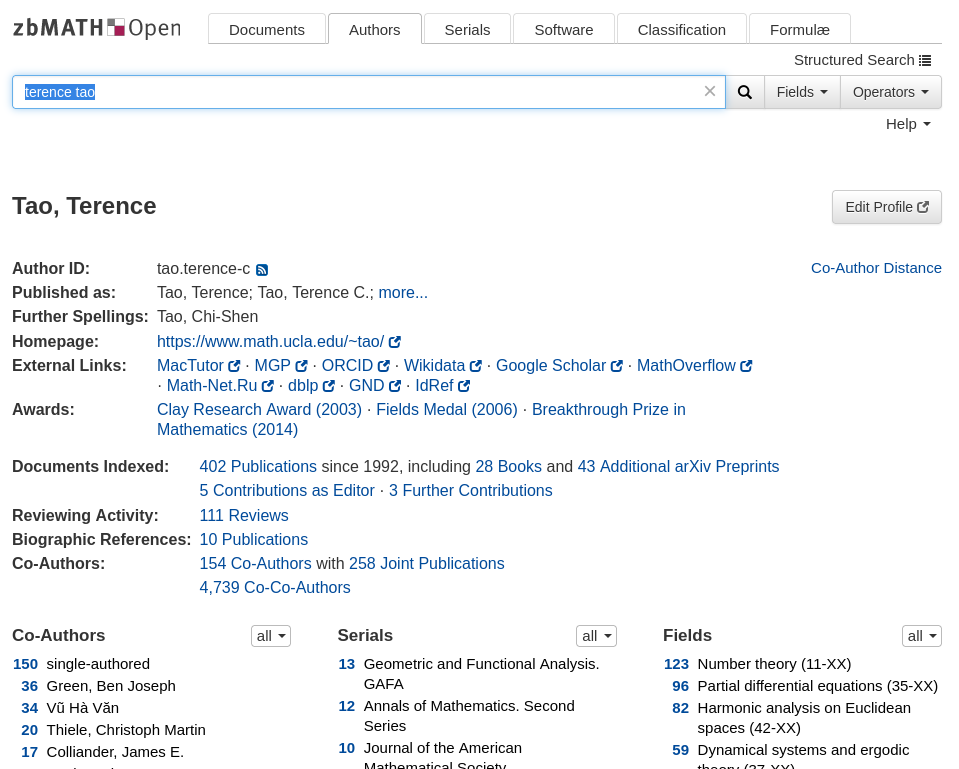}
\captionof{figure}{zbMATH author profiles with additional arXiv preprints}
\end{center}

Another aspect is that, due to this addition, the share of Open Access documents available in zbMATH Open inceases significantly, and approaches now about 1 million (i.e., \(>20\)\%). Since these arXiv metadata are also included into the zbMATH Open APIs \cite{ST21, FM23}, this creates additional opportunities for further research and infrastructure projects based on an open mathematics corpus.

\section{Technical challenges and limitations}\label{tech}

\subsection{Matching}\label{matching}

At a first glance it seems relatively easy to find a corresponding published version for an arXiv preprint. 
All one needs to do is to search for a document in zbMATH Open with the same title and author(s). Indeed, this approach leads to a lot of correct assignments. 
However, this will also miss a lot of matches as there is often a difference between the (latest) arXiv version and the published article indexed at zbMATH Open. 
The discrepancies include: different layout, numbering of theorems and sections, addition or deletion of content, or change of the title or abstract. 
Even the authors can differ in the published version. Thus, this naive matching approach would lead to a lot of duplicate entries after the arXiv integration. 
To avoid this problem, a precise matching procedure is needed. Our new approach is based on the DOI, title, authors and abstract of an arXiv article. 
It turns out that it has a higher accuracy as the naive comparison of title and authors.

Our matching algorithm has two steps. First, if an arXiv article has a DOI of a related published version given, then we search for an article with this DOI at zbMATH. If a unique article is found, we stop and return this article. Otherwise, a second matching step is executed, which is based on the title, author and abstract similarity. The idea behind this approach is that it might be that one of those three metadata information differs in the published version from the preprint, however, it is unlikely that all three differ simultaneously in a substantial manner. 

In more detail, the second matching step works as follows. The title and authors of an arXiv preprint are used to search for a variable number of ``candidate" zbMATH articles that might or might not match the input preprint. For every candidate a three dimensional vector is computed that contains the similarity of the title, authors and abstract of the candidate article to the input. Finally, a random forest classifier is used to decide whether a candidate matches or not. If more than one candidate matches, then the one with the lexicographic smallest similarity vector is returned (the similarities are scaled to $[0,1]$ and a smaller number means higher similarity). 

A disadvantage of our new approach, as compared  to the naive one, is its higher complexity and that training data is needed to train the random forest classifier. The training data consists of pairs of correct arXiv, zbMATH matches which were generated using DOI-Matching. This leads to some wrong pairs in the training data as the DOI given by arXiv or by zbMATH might be incorrect. Overall, the DOI information is very precise, so the training data is good enough to generate a random forest with a good performance. On some test data, which was also generated using DOI-Matching, the classifier matching gives a precision of 99.51 \% and a recall of 96.89 \%.

Does our matching approach also makes a difference in practice? In total our new two step matching algorithm found 250,425 matches (14th December 2023). 73,567 of those come from the DOI matching (first step) and 176,858 from the classifier matching (second step). From the 176,858 classifier matches, 144,825 have exactly the same title and authors (after some normalization steps). Which means that our approach gives us 32,033 new matches compared to a naive comparison of title and authors.

\subsection{Author disambiguation}\label{authors}

The release of the zbMATH Open interface with the additional arXiv entries is also internally coupled with a new version of the automated author disambiguation. 
Beside slightly improved matching results, its main advantage is mostly invisible to the outside -- its much more modularized structure was a prerequisite to handle both the new amount and somewhat different structure of the newly added items. Naturally, the assignments of arXiv preprints are somewhat less reliable than for the published corpus, due the a higher occurrence of metadata errors (e.g., in separating author names). Also, the unstable nature of arXiv entries must be handled appropriately in the assignment tables. Moreover, there is also the option that arXiv papers may be withdrawn -- currently, we follow the arXiv policy that they are nevertheless still available.

\begin{center}
\includegraphics[width=.5\textwidth]{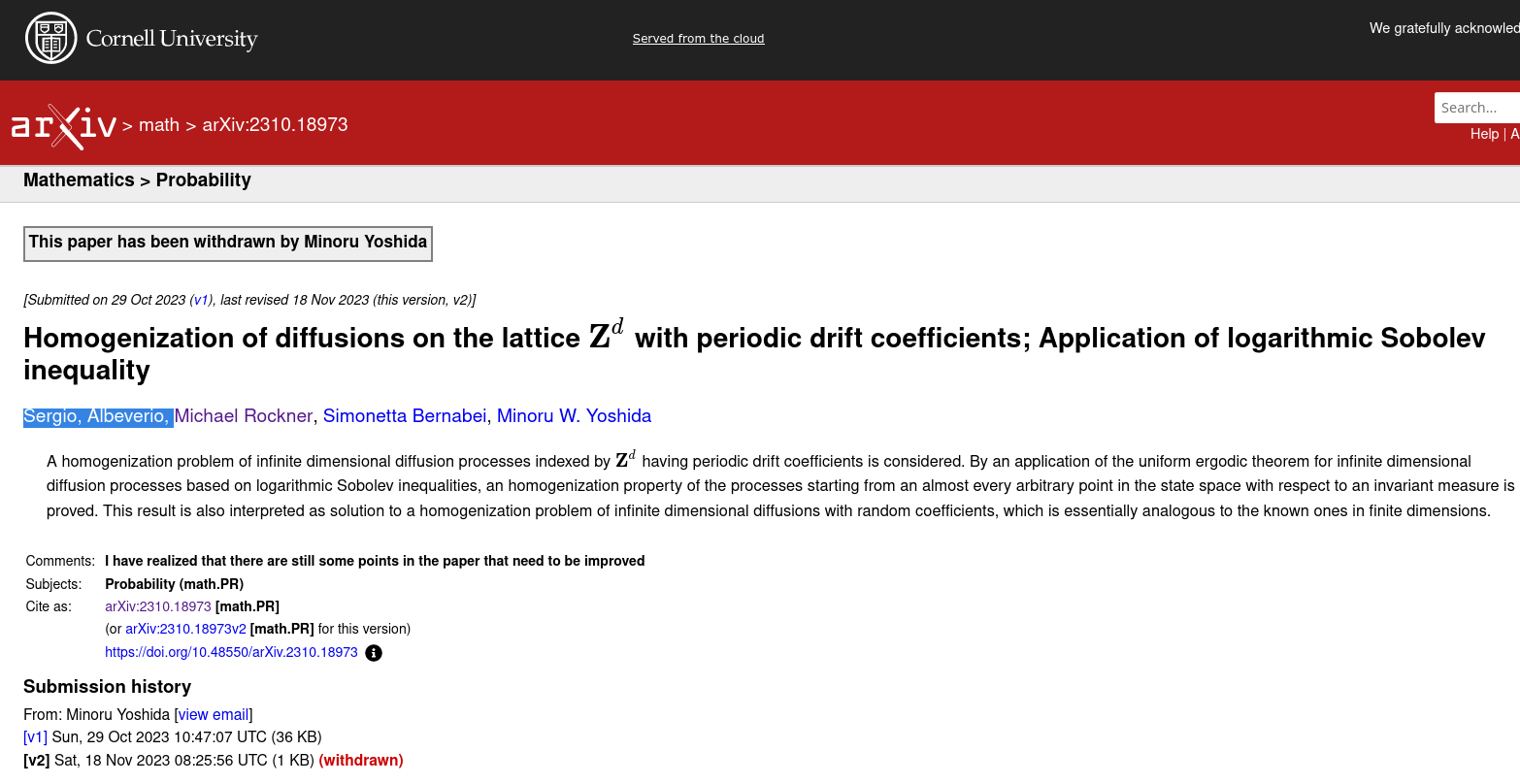}
\captionof{figure}{A withdrawn arXiv submission with a separator error}
\end{center}

Perhaps most importantly, many new profiles for mostly young researchers have been created who have currently only arXiv submissions, but not yet published articles. Their enhanced visibility will 
be one of the most immediate effects of the integration.

\begin{center}
\includegraphics[width=.5\textwidth]{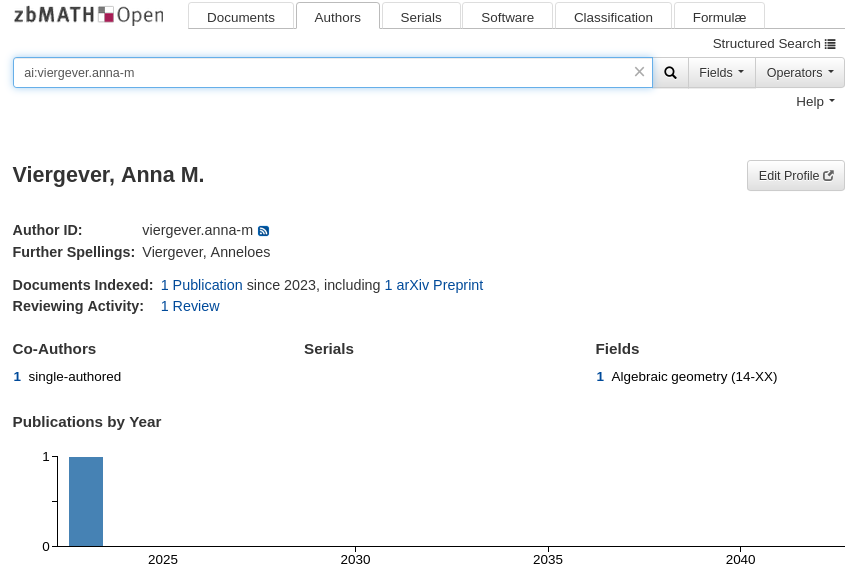}
\captionof{figure}{A new author profile based on arXiv submissions}
\end{center}

\section{Future challenges}
\label{sec:future}

As discussed in Section \ref{features}, the additional arXiv entries are partially integrated into the whole zbMATH Open framework, but not  all features are currently available.
The addition of an automated preliminary classification for the items which have not yet an MSC assigned by the authors is technically feasible and would not be very difficult to incorporate. 
Also, for some arXiv submissions we already collect software information, although it is not yet displayed. Once fully available, this would further allow for the integration of these entries into swMATH. Likewise, the interlinking with OEIS, DLMF, or MathOverflow -- as provided for stable zbMATH Open documents -- is not yet implemented, and  requires further development work. 
Probably the  most challenging task will be the integration into the zbMATH citation database, due to the various formats and unstable version entries of arXiv references.

On the positive side, the availability of open \LaTeX{} sources for arXiv documents opens up a considerable future potential. 
Depending on the license, a multitude of full-text features could be developed and implemented, ranging from full-text formula search to features involving large language models based on a reliable mathematical corpus.   
With the information available on the zbMATH Open APIs, we invite the community to develop interesting features which include the newly added information!

\end{document}